\documentclass[12pt]{article}

\usepackage[normalem]{ulem}
\usepackage[mathscr]{eucal}
\usepackage[version=3]{mhchem}
\usepackage{stmaryrd}
\usepackage{stackrel}

\usepackage{graphicx}

\usepackage{epsfig,latexsym,amsfonts,amsmath,amsthm,amssymb,amsbsy,multirow,slashed,wasysym,textcomp,dsfont,comment,mathtools,cancel,cite,datetime,appendix,BOONDOX-calo}
\usepackage{tocloft}
\usepackage{bbold}
\usepackage{tikz}
\usetikzlibrary{decorations.pathreplacing,calligraphy}
\usetikzlibrary{backgrounds}
\usetikzlibrary{patterns}
\usetikzlibrary{arrows.meta}
\tikzstyle{bag} = [align=center]
\usetikzlibrary{decorations.pathmorphing}
\def\bea{\begin{eqnarray}}
\def\eea{\end{eqnarray}}

\usepackage{hyperref}
\usepackage{subcaption}

\usepackage{array}
\usepackage{booktabs}

\newcommand{\badat}{\begin{alignedat}}
\newcommand{\eadat}{\end{alignedat}}

\def\be{\begin{equation}}
\def\ee{\end{equation}}
\def\ba{\begin{aligned}}
\def\ea{\end{aligned}}

\usepackage{xcolor}

\newcommand{\pink}[1]{\textcolor{\pink}{#1}}

\definecolor{dblue}{rgb}{0.2,0.50,0.80}

\def\D{\hat{\mathcal{D}}}

\usepackage{hyperref}

\DeclareSymbolFont{extraup}{U}{zavm}{m}{n}
\DeclareMathSymbol{\varheart}{\mathalpha}{extraup}{86}
\DeclareMathSymbol{\vardiamond}{\mathalpha}{extraup}{87}

\DeclareFontFamily{OT1}{pzc}{}
\DeclareFontShape{OT1}{pzc}{m}{it}{<-> s * [1.10] pzcmi7t}{}
\DeclareMathAlphabet{\mathpzc}{OT1}{pzc}{m}{it}

\definecolor{vert}{rgb}{0.1367 0.543 0.1367}

 \newcommand\omicron{o}

\PassOptionsToPackage{linecolor=blue,backgroundcolor=blue!25,bordercolor=blue,textsize=scriptsize}{todonotes}
\usepackage{todonotes}

\numberwithin{equation}{section} 

\begin{document}

 \begin{titlepage}
  \thispagestyle{empty}
  \begin{flushright}
  \end{flushright}
  \bigskip
  \begin{center}

        \baselineskip=13pt {\LARGE {
        Gravity From a Color Symmetry II:  \\[.5em]
        Celestial Color Kinematics for Mass and Spin
       }}
     
      \vskip1cm 

   \centerline{ 
   {Alfredo Guevara}\footnote{aguevaragonzalez@ias.edu} ${}^{\clubsuit,{}\spadesuit{},{}\vardiamond}$ }

\bigskip\bigskip

\centerline{\em${}^\clubsuit$ 
\it School of Natural Sciences, Institute for Advanced Study, Princeton, NJ 08540 USA
}

\bigskip

\centerline{\em${}^\spadesuit$ 
\it Center for the Fundamental Laws of Nature, Harvard University, Cambridge, MA 02138 USA
}

\bigskip

\centerline{\em${}^\vardiamond$ 
\it Society of Fellows, Harvard University, Cambridge, MA 02138 USA
}

\bigskip\bigskip

\end{center}

\begin{abstract}
A realization of gravitational
amplitudes based in the large $N$ limit of a certain 2d $SU(N)$ Kac-Moody theory has been recently proposed.
We relate this proposal to Color Kinematics (CK) duality and present an extension to EFT
amplitudes for matter particles with any mass and spin. In particular, we recast these EFT amplitudes as celestial correlation functions and show they posses a chiral $w_{1+\infty}$
symmetry algebra if they are minimally coupled in the bulk. Massive states lead to an off-shell 1-parameter deformation of the algebra. Finally, we argue that in the limit $S\to\infty$ these states correspond to the
Kerr black hole and we rediscover a classical $w_{1+\infty}$ action
of Penrose. 
\end{abstract}

\end{titlepage}

\section{Introduction}
A novel symmetry of gravitational scattering amplitudes
has been pinpointed through the celestial holography program \cite{Guevara:2021abz,Strominger:2021mtt}, 
by recasting the gravitational S-Matrix in terms of correlations functions
of a two-dimensional Celestial CFT (CCFT). The symmetry,
dubbed $Lw_{1+\infty}$ or $w_{1+\infty}$ for simplicity, has corresponding Ward identities that can be matched to exponentiated soft theorems in the S-Matrix \cite{Guevara:2019ypd} language. 

Even though
this formulation was presented in a boost eigenstate basis, appropriate
to the CCFT, it has become clear that a tantamount statement
can be made in the 4d momentum S-Matrix, see e.g. \cite{Guevara:2022qnm,Costello:2022wso}, and that the symmetry leads to profound consequences in a putatitive holographic 2d-4d duality \cite{Costello:2022wso}. In particular, the duality enjoys description in twistor space \cite{Costello:2023prl} which makes explicit the $w_{1+\infty}$ symmetry \cite{Adamo:2022wso,Monteiro:2022lwm} as well as its 4d anomaly cancellations \cite{Doran:2024abc,Bittleston:2022abc}.

In \cite{Guevara:2022qnm} we proposed that the holographic theory enjoys
the $N\to\infty$ limit of a particular $U(N)$ Kac-Moody current
algebra, which can be identified with the loop-wedge of $w_{1+\infty}$. Its two-dimensional Jacobi
identity, a consistency condition of the OPE, is equivalent to
the kinematic Jacobi identity of 4d perturbative scattering amplitude of gravitons \cite{Mago:2021wje}.
However, the kinematic Jacobi identity of the S-Matrix is the main
avatar of the so-called color kinematics duality \cite{Bern:2019prr}, which relates a
sheer diversity of gravitational amplitudes to their gauge theory
analogs. In particular, it holds for a large number of EFT operators
at tree-level including massless or massive particles \cite{Bautista:2019evw,Johansson:2019dnu}. In the
spirit of this, it is expected that the 2d Jacobi identity of $w_{1+\infty}$
is realized also in a much more general scenario. Moreover, if this
is so, it is expected that a 2d holographic picture should be present
for such EFT amplitudes.

In this paper we show that this is indeed the case, for any particle
spectrum compounded of massless or massive particles of any spin,
as long as the following requirement holds: The coupling of a positive-helicity
(+) graviton to any other particle (including negative-helicity gravitons)
must be minimal, in the sense of \cite{Arkani-Hamed:2017jhn}. In such cases, we will see
that
\begin{enumerate}

\item A 2d color symmetry is realized in the scattering amplitude. Its
associativity can be derived from the 4-pt amplitude (along the lines of e.g. \cite{Ball:2022vzc,Mago:2023abc,Guevara:2022qnm}).

\item A notion of color kinematics emerges, in particular we can define
numerators $n_{i}(\epsilon,k)$ that satisfy the kinematic Jacobi
identity,
\end{enumerate}
\begin{equation}
    n_S+n_T+n_U=0\,.
\end{equation}
These two facts stem from remarkable properties of QFT amplitudes
for spin-$S$ particles that were mainly outlined in \cite{Bautista:2019tdr}.
To understand how this comes about, consider a toy version of the
above situation arising in the context of QED/QCD. For a (+) photon
of momentum $k$ and polarization $\epsilon$ coupled to a matter
particle of momentum $p_{1}$ and spin $S=1$, the amplitude develops
a singularity as $p_1\cdot k/E^{2}\to0$, and the corresponding factorization
can be expressed in terms of two EFT form factors:

\begin{equation}
\mathcal{A}_{n+1}(\{\epsilon,k\},\{\varepsilon_{1},p_{1}\},\ldots)\to q\frac{(\epsilon\cdot p_{1})}{p_1\cdot k}\langle\varepsilon_{1}|\left(\mathbb{I}+g\frac{F_{\mu\nu}J^{\mu\nu}}{2\epsilon\cdot p_{1}}\right)|\varepsilon_{I}\rangle\mathcal{A}_{n}(\{\varepsilon^{I},p_{I}=p_{1}+k\},\ldots)\label{eq:qcdex}
\end{equation}
where we sum over internal states $|\varepsilon_{I}\rangle$. Recall
that $F_{\mu\nu}=2k_{[\mu}\epsilon_{\nu]}$ and that $J_{\mu\nu}$
is the $S=1$ Lorentz generator on $|\varepsilon_{I}\rangle$, hence
$g$ is the standard gyromagnetic ratio. If $p_1$ is massless this is a photon or gluon splitting function and $g$ is fixed to $g=2$. In the massive case, the
`minimal coupling EFT' is defined by a smooth massless limit \cite{Arkani-Hamed:2017jhn},
thus implicitly imposing $g=2$. 

\begin{figure}[h]
\centering
\includegraphics[width=0.9\textwidth]{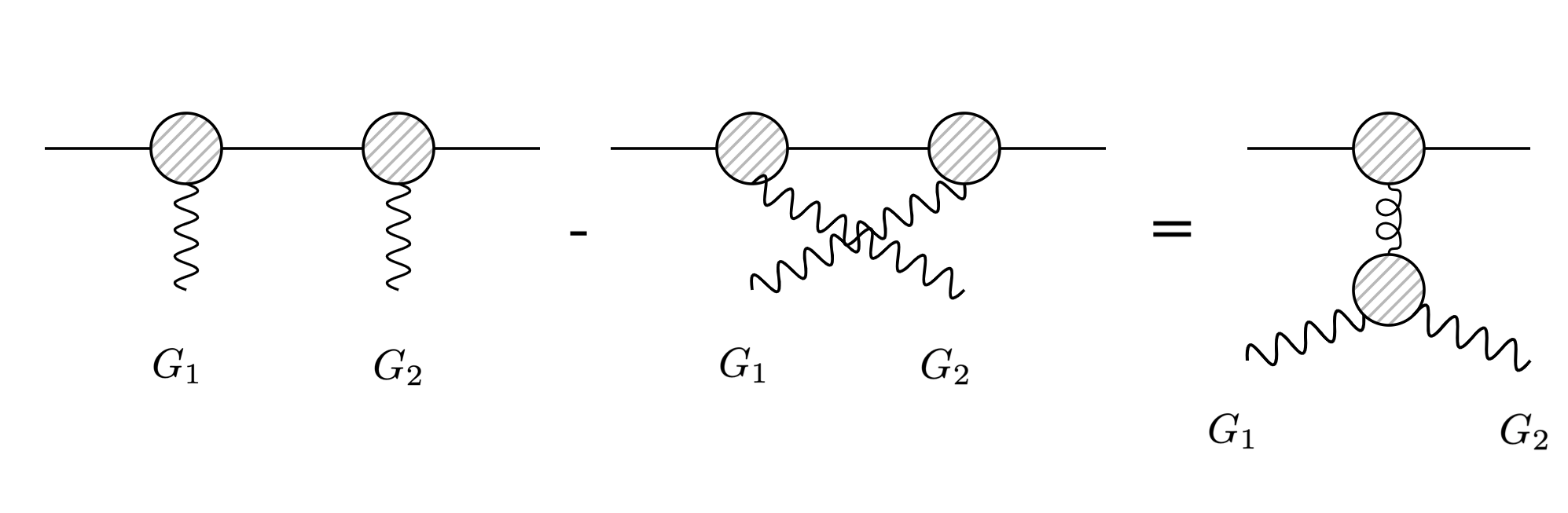}
\caption{Graphic depiction of the commutation relation $G_1G_2 - G_2 G_1 = [G_1,G_2]$ that realizes the kinematical $w_{1+\infty}$ algebra. The solid line represents a matter particle.}\label{fig:figo}
\end{figure}
The key point that connects to a color realization is the following:
On the support of the singularity $p_{1}\cdot k=0$ one finds that
the momenta, being spin-1 vectors themselves are related through a
particular Lorentz transformation:
\begin{equation}
p_{I}=\exp\left(\frac{F_{\mu\nu}J^{\mu\nu}}{2\epsilon\cdot p_{1}}\right)p_{1}    \,.
\end{equation}
It can be shown, using the on-shell condition, that the exponential can be expanded to truncate at linear order. Remarkably, it is
only in the case $g=2$ where the polarization vector $|\varepsilon_{I}\rangle$
in (\ref{eq:qcdex}) undergoes precisely the same transformation.
Hence, we can combine orbital and intrinsic parts into the operator-valued
identity
\begin{equation}
\mathcal{A}_{n+1}^{\textrm{QCD}}(\{\epsilon,k\},\{\varepsilon_{1},p_{1}\},\ldots)\to q\frac{(\epsilon\cdot p_{1})}{p_1\cdot k}\exp\left(\frac{F_{\mu\nu}J^{\mu\nu}}{2\epsilon\cdot p_{1}}\right)\mathcal{A}_{n}^{\textrm{QCD}}(\{\varepsilon_{1},p_{1}\},\ldots)\,,\label{eq:etxps}
\end{equation}
where the operator acts on the momentum state $\{\varepsilon_1,p_1\}$. Analogously, for gravity we define the minimal coupling by the relation
\begin{equation}
\mathcal{A}_{n+1}^{\textrm{GR}}(\{\epsilon,k\},\{\varepsilon_{1},p_{1}\},\ldots)\to\kappa\frac{(\epsilon\cdot p_{1})^{2}}{p_1\cdot k}\exp\left(\frac{F_{\mu\nu}J^{\mu\nu}}{2\epsilon\cdot p_{1}}\right)\mathcal{A}_{n}^{\textrm{GR}}(\{\varepsilon_{1},p_{1}\},\ldots)\,.\label{eq:grf}
\end{equation}
This is the content of the color symmetry proposed in \cite{Guevara:2022qnm}:
In particular, it was observed there that for a massless particle
with $p_{1}=\lambda_{1}\tilde{\lambda}_{1}$ and any spin, the colinear
limits are controlled by 
\begin{equation}
\mathcal{A}_{n+1}^{\textrm{GR}}(\{\epsilon,k\},\{\varepsilon_{1},p_{1}\},\ldots)\to\frac{[\tilde{\eta}\tilde{\lambda}]}{\langle\eta\lambda\rangle}\exp\left(\tilde{\eta}^{\dot{\alpha}}\frac{\partial}{\partial\tilde{\lambda}^{\dot{\alpha}}}\right)\mathcal{A}_{n}^{\textrm{GR}}(\{\varepsilon_{1},p_{1}\},\ldots)\label{eq:efsa}
\end{equation}
where $k=\eta\tilde{\eta}$ is the momentum in bi-spinor notation. Remarkably, the spinor $\tilde{\eta}^{\dot{\alpha}}$ can be
interpreted as an adjoint index of a $U(N\to\infty)$ color symmetry, which is isomorphic to
$w_{1+\infty}$. It shall become clear throughout this work that (\ref{eq:etxps})
and (\ref{eq:efsa}) reflect precisely the same color-like structure, and moreover
(\ref{eq:etxps}) can be extended to massive particles of arbitrary spin in a minimal
coupling EFT. 

The exponential operator in (\ref{eq:etxps}) is nothing but the structure
constant of a representation of $w_{1+\infty}$ for arbitrary spin.
The main result of section \ref{eq:offshells} is the implementation of the Baker-Campbell-Haussdorf
formula to show that such operators satisfy a Jacobi identity. The identity emerges when inserting two graviton states $G_1,G_2$, each generating a color transformation. It can be written in a diagramatic way as in
Figure \ref{fig:figo}. For the momentum space S-matrix, this is nothing but the kinematic Jacobi algebra, at
the root of the color-kinematics duality. 

One of the goals of the celestial program is to define the duality in curved backgrounds \cite{Bittleston:2023abc,Costello:2023bmi,Bittleston:2024rqe,Bogna:2024abc}. On the other hand, massive spinning amplitudes, and in particular their color-kinematics duality, have been fruitfully applied to describe certain black hole spacetimes \cite{Monteiro:2014cda,Guevara:2021abx}. Here, we will argue that certain type D spacetimes such as the Kerr metric can emerge in this form from the `color structure constants' \eqref{eq:grf} when interpreted in twistor space via the so-called Penrose transform. 

In the rest of the paper we outline the details of this construction. We start off by introducing spinor kinematics and setting up notation in section \ref{sec:kin}. In section \ref{sec:color} we will formulate the Jacobi identity both as on-shell 2d celestial associativity as well as check it explicitly in massive amplitudes. In section \ref{eq:offshells} we provide an off-shell version which involves a 1-parameter `massive' deformation of $w_{1+\infty}$. In section \ref{eq:spact} we flesh out the application of the color structure to derive classical spacetimes. We wrap up with a discussion of future directions.

\section{Kinematics: Massive Spinor Representation}\label{sec:kin}

We start this section by introducing notation and setting up the momentum basis in terms of spinor variables. We find this basis more suitable to describe the color structure, but a conformal/Mellin transformed basis is also possible (see discussion).

In \cite{Guevara:2022qnm} a color structure for the massless matter (hard) particles
was unveiled by introducing spinors
\begin{align}
|\lambda\rangle & =(1,z_{1})=|+\rangle+z_{1}|-\rangle\,\,.\nonumber \\
 & |\lambda]=(\tilde{\lambda}_{1},\tilde{\lambda}_{2})\label{eq:mlsp}
\end{align}
where we have fixed a chiral basis $|\pm\rangle$ in order to introduce
the inhomogeneous coordinate $z_{1}\in\mathbb{CP}^{1}$. The antichiral
component is interpred as a $SU(N)$ adjoint index, if restricted
to the lattice $|\lambda]\in Z_{N}\times Z_{N}$. Then, in the large
color limit $N\to\infty$, the wedge of the algebra $w_{1+\infty}$ is obtained as a continuum
limit. The 4D momentum of a massless particle is then $p^{\mu}\sigma_{\mu}=|\lambda\rangle[\tilde{\lambda}|$. 

For a massive particle, we will extend the above spinor variables along the lines of \cite{Arkani-Hamed:2017jhn},
\begin{equation}
p_{1}=p_{1}^{\mu}\sigma_{\mu}=|1_{+}\rangle[1_{-}|-|1_{-}\rangle[1_{+}|\,,\quad p_1^2\neq 0.
\end{equation}
The indices $a=\pm$ are $SU(2)$ indices reflecting the little group
redundancy $|1_{a}\rangle=U_{a}^{b}|1'_{b}\rangle$,$|1_{a}]=U_{a}^{b}|1'_{b}]$
of $p_{1}$. Note $U_{a}^{b}$ has 3 indepedent components. Furthermore,
there is an extra scale transformation $|1_{a}\rangle\to t|1_{a}\rangle\:|1_{a}]\to t^{-1}|1_{a}]$,
for a total of four redundancies (forming a $GL(2)$ algebra). These can be used to fix four components.
For the purpose of manifesting the (antichiral) color symmetry in
$|1_{a}]$ we will choose
\begin{equation}
p_{1}=|+\rangle[\lambda|+|-\rangle[\mu|=\left(\begin{array}{cc}
\lambda_{+} & \mu_{+}\\
\lambda_{-} & \mu_{-}
\end{array}\right) \label{eq:p1fix}
\end{equation}
such that $\langle-+\rangle=1$ and $[\mu\lambda]=p_{1}^{2}$. Then
$|\lambda]$ and $|\mu]$ will play the role of color indices, the
former being analogous to the massless case \eqref{eq:mlsp}. For a spin
$S$ particle, it is known that polarization tensors take the $2S+1$
configurations
\begin{equation}
\epsilon_{\alpha_{1}\cdots\alpha_{k}\alpha_{k+1}\ldots\alpha_{2S}}^{(k)}=\frac{1}{m^{2S}}\tilde{\lambda}_{(\dot{\alpha}_{1}}\ldots\tilde{\lambda}_{\dot{\alpha}_{k}}\tilde{\mu}_{\dot{\alpha}_{k+1}}\ldots\tilde{\mu}_{\dot{\alpha}_{2S})}\,,\,\,\,0\leq k\leq2S\,.\label{eq:polt}
\end{equation}
For a given spinor $|\lambda]$, it may be convenient to introduce
a reference $|r_{\lambda}]$ so that $[r_{\lambda}\lambda]=m^{2}$
and $r_{\lambda}\to0$ in the massless limit. Using it, we can associate
a coordinate $Z_{1}\in\mathbb{CP}^{1}$ to the massive particle, such
that 
\begin{equation}
    |\mu]=Z|\lambda]+|r_{\lambda}]  \label{eq:pffm}
\end{equation}
One can check that the massless limit becomes $p_{1}\to|\lambda\rangle[\lambda|$,
with $|\lambda\rangle=(1\,,\,Z)$ as expected. 

For the massless graviton we will take momenta $k$ and polarization
$\epsilon_{\mu\nu}=\epsilon_{\mu}\epsilon_{\nu}$ where
\begin{align}
k & =|\eta\rangle[\eta|\,\,\,,|\eta\rangle=|+\rangle+z|-\rangle\,,\nonumber\\
\epsilon & =\partial_{z}k=|-\rangle[\eta|
\end{align}
Introducing an inhomogeneous coordinate $z\in\mathbb{CP}{}^{1}$ positons
the graviton in the complexified celestial sphere, or more precisely in the celestial torus if we assume Kleinian signature. The price to pay is fixing the little-group of the graviton \cite{Guevara:2022qnm}, which can be restored by appropriate factors of $\langle\eta r\rangle$
for a reference $r$. 

We will now flesh out the advantage of the above parametrizations. First,
note that
\begin{align}
2\epsilon\cdot p_{1} & =[\lambda\eta]\,,\nonumber \\
2k\cdot p_{1} & =[\lambda\eta]\left(z-\frac{[\mu\eta]}{[\lambda\eta]}\right)=[\lambda\eta]\left(z-Z_{1}-\frac{[r_{\lambda}\eta]}{[\lambda\eta]}\right)\,,\label{eq:kpz}
\end{align}
which reduce to the massless case when $|r_{\lambda}]\to0$.
This will be important to connect to the color structure. Second,
using

\begin{equation}
-\frac{\partial}{\partial p_{1}^{\alpha\dot{\alpha}}}=|-\rangle_{\alpha}\frac{\partial}{\partial\lambda^{\dot{\alpha}}}-|+\rangle_{\alpha}\frac{\partial}{\partial\mu^{\dot{\alpha}}}\,,\label{eq:dp2}
\end{equation}
one finds 
\begin{align}
\mathcal{D}_{\eta}:=-2k^{\mu}\frac{\partial}{\partial p^{\mu}} & =[\eta\frac{\partial}{\partial\lambda}]+z[\eta\frac{\partial}{\partial\mu}]\,.\label{eq:kdp}
\end{align}
This displacement generator $\mathcal{D}_{\eta}$ is closely related
to the angular momentum. Let us argue why this is the case. Consider
first the scalar case $S=0$, for which there is no polarization tensors:
This means that the total angular momentum is composed only of the
`orbital' piece, namely $J_{\mu\nu}=2p_{[\mu}\frac{\partial}{\partial p^{\nu]}}$.
We can further extract the chiral part of the generator via $J_{\dot{\alpha}\dot{\beta}}:=\frac{1}{2}J_{\mu\nu}\sigma_{\dot{\alpha}\dot{\beta}}^{\mu\nu}$.
After a short computation, using (\ref{eq:dp2}), it becomes

\[
J_{\dot{\alpha}\dot{\beta}}=\tilde{\lambda}_{(\dot{\alpha}}\frac{\partial}{\partial\tilde{\lambda}^{\dot{\beta})}}+\tilde{\mu}_{(\dot{\alpha}}\frac{\partial}{\partial\tilde{\mu}^{\dot{\beta})}}\,.
\]
But this is crearly the action of the angular momentum on a generic
spin-$S$ representation, if the polarization tensors are also constructed
from $|\lambda]$ and $|\mu]$ as in (\ref{eq:polt}). Now let us further introduce, as usual, the field-strength $F_{\mu\nu}^{+}=2k_{[\mu}\epsilon_{\nu]}^{+}$.
It is a chiral tensor in the sense that $F_{\dot{\alpha}\dot{\beta}}=\frac{1}{2}F_{\mu\nu}\sigma_{\dot{\alpha}\dot{\beta}}^{\mu\nu}=\eta_{\dot{\alpha}}\eta_{\dot{\beta}}$
and $F_{\alpha\beta}=\frac{1}{2}F_{\mu\nu}\sigma_{\alpha\beta}^{\mu\nu}=0$.
With this two ingredients we define
\begin{equation}
\mathcal{\hat{D}}_{\eta}=\frac{F_{\mu\nu}^{+}J^{\mu\nu}}{\epsilon^{+}\cdot p_{1}}=\frac{F_{\dot{\alpha}\dot{\beta}}J^{\dot{\alpha}\dot{\beta}}}{[\lambda\eta]}=[\eta\frac{\partial}{\partial\lambda}]+\frac{[\mu\eta]}{[\lambda\eta]}[\eta\frac{\partial}{\partial\mu}]\,.\label{eq:fj}
\end{equation}
The antichiral generators satisfy 
\begin{align}
    [J_{\dot{\alpha}_1\dot{\beta}_1},J_{\dot{\alpha}_2\dot{\beta}_2}]&= \epsilon_{\dot{\alpha}_2(\dot{\alpha}_1} J_{\dot{\beta}_1)\dot{\beta}_2} + \epsilon_{\dot{\beta}_2(\dot{\alpha}_1} J_{\dot{\beta}_1)\dot{\alpha}_2} \\
    [J_{\dot{\alpha}\dot{\beta}},\lambda_{\dot{\gamma}}]&= \epsilon_{\dot{\gamma} \dot{(\alpha}} \lambda_{\dot{\beta})} \label{eq:prealg}
\end{align}
As mentioned, the generator $\mathcal{\hat{D}}_{\eta}$ is one of
the main objects in this work. Note that on the support of $k\cdot p_{1}\to0$
we have from (\ref{eq:kpz}) that $z_{1}\to\frac{[\mu\eta]}{[\lambda\eta]}$,
which in turn implies

\[
\mathcal{\hat{D}}_{\eta}\to\mathcal{D}_{\eta}
\]
As $k\cdot p_{1}=0$ is an on-shell condition for the massive particle,
we refer to $\mathcal{D}_{\eta}$ as `off-shell' and to $\hat{\mathcal{D}}_{\eta}$
as `on-shell'. Indeed, $\hat{\mathcal{D}}_{\eta}$ being a Lorentz transformation, preserves the on-shell condition $[\lambda \mu]=m$. Importantly, the off-shell generator (\ref{eq:kdp})
is linear in $\eta_{\dot{\alpha}}$. This mimics what happens in the massless limit,
where $\mathcal{\hat{D}}_{\eta},\mathcal{D}_{\eta}$ correspond
to $[\eta\frac{\partial}{\partial\lambda}]$ \cite{Guevara:2022qnm}. We will see in section \ref{eq:offshells}
that in the on-shell or massive case, the non-linear terms in $\eta$
will incorporate corrections to the $w_{1+\infty}$ algebra. 

\subsection{Dictionary for massive states}

Using the above parametrization, and following the aforementioned momentum space
construction, we will reinterpret graviton and matter particles
in the S-Matrix as currents and primaries in a CFT. This relies heavily on Klein signature since it allows us to define collinear singularities for massive states, which is not possible in the Lorentzian case.

More precisely,
a graviton state with momenta $k=\eta\tilde{\eta}$ is denoted by
the insertion
\[
a^{+2}(\eta,\tilde{\eta})\leftrightarrow G^{\tilde{\eta}}(z)
\]
They satisfy the $Lw_{1+\infty}$ algebra in the form of the Kac-Moody OPE (we set $\kappa=1$)
\begin{equation}
      G^{\eta_{1}}(z_{1})G^{\eta_{2}}(z_{2}) \sim \frac{[12]}{z_1-z_2}    G^{\eta_1+\eta_{2}}(z_{2})\label{eq:womf}
\end{equation}
We recall that the wedge algebra appearing in e.g. \cite{Strominger:2021mtt} is obtained by writing $|\eta] = \omega (1\,\,,\,\,\bar{z})$ and expanding both sides in $\omega_1,\omega_2.$

A matter particle of any spin and momentum $p_{1}$ will be
denoted by
\[
a_{S}(p_{1})\leftrightarrow O_{S}^{\tilde{\lambda}}(z_{1})=O_{S}^{\tilde{\lambda},\tilde{\mu}}
\]
Using this dictionary we are able to place the massive state at a puncture $z_{1}$ that parametrizes its spinor $\mu$ via \eqref{eq:pffm}. Recall we will keep only the \textit{left action} of the Lorentz group $SL(2,\mathbb{R})_L$ (acting on $z_i$) since the right action is absorbed into the color structure.

We will then study the connected correlation functions of these puncture insertions, which are mapped
to the scattering amplitude via a celestial dictionary 
\begin{equation}
\langle G^{\tilde{\eta}_{1}}(z_{1})G^{\tilde{\eta}_{2}}(z_{2})\cdots O_{S}^{\tilde{\lambda}}(z)O_{S'}^{\tilde{\lambda}'}(z')\cdots\rangle=\mathcal{M}\left(\{\epsilon_{1},k_{1}\}^{+2},\{\epsilon_{2},k_{2}\}^{+2}\cdots\{\varepsilon,p\}^{S},\{\varepsilon',p'\}^{S'}\cdots\right) \label{eq:dcin}
\end{equation}
In this work we consider dressed amplitudes $\mathcal{M}(\cdots)$
which include the momentum conservation delta function. If $k_{i}$
and $p_{j}$ are massless and massive momenta respectively, momentum
conservation is the support
\begin{equation}
\delta^{4}(\sum_{i}k_{i}+\sum_{j}p_{j})=\delta^{2}(\sum_{i}\tilde{\eta}_{i}+\sum_{j}\tilde{\lambda}_{j})\delta^{2}(\sum_{i}z_{i}\tilde{\eta}_{i}+\sum_{j}\tilde{\mu}_{j})
\end{equation}
The first factor represents color conservation and will enter in the color kinematics duality. In the next section we reinterpret
the second factor as a meromorphic function with poles in $z_{i}$.
Additionally, we will use the standard two-point connected function for massive
spin $S$ states
\begin{align}
\langle O_{S}^{\tilde{\lambda}}(z)O_{S'}^{\tilde{\lambda}'}(z')\rangle & =\langle\varepsilon^{(k)}|\varepsilon'^{(j)}\rangle\,\delta^{2}(\lambda+\lambda')\delta\left(\frac{[\mu\xi]}{[\lambda\xi]}-\frac{[\mu'\xi]}{[\lambda'\xi]}\right)\nonumber\\
 & \propto\delta^{k+j-2S}\delta^{2}(\lambda+\lambda')\delta(Z-Z')\,,\label{eq:2ptss}
\end{align}
where $\xi$ is any reference spinor and 
\begin{equation}    \langle\varepsilon^{(k)}|\varepsilon'^{(j)}\rangle=\epsilon_{\alpha_{1}\ldots\alpha_{2S}}^{(k)}\epsilon^{'(j)\alpha_{1}\cdots\alpha_{2S}}\,,
\end{equation}
which are given by (\ref{eq:polt}). In general this factor will transform
covariantly under $SU(2)$ acting on either particle; however, due
to our particular little-group fixing (\ref{eq:p1fix}) together with
$[\lambda\mu]=[\lambda'\mu']$ we have $\langle\varepsilon^{(k)}|\varepsilon'^{(j)}\rangle\propto\delta^{k+j-2S}$.
The massless limit of this two-point function is obtained with $k=0,j=2h$
and thus $\langle\varepsilon^{(k)}|\varepsilon'^{(j)}\rangle\to1$. 

Another implication of the above is the following formula for the LIPS (Lorentz Invariant Phase Space), namely the integral measure conjugate to \eqref{eq:2ptss}
\begin{equation}
    \sum_k \int d^2\lambda \, dz \, \langle O_{S}^{\lambda}(z)O_{S'}^{\lambda'}(z')\rangle = 1\,.
\end{equation}
One can check that the measure $d^2\lambda \, dz$ is proportional to the standard dLIPS in $(2,2)$ signature, $d^4p \,\delta(p^2-m^2)$, under the parametrization \eqref{eq:p1fix}.

\section{Color-Kinematics and On-shell $w_{1+\infty}$}\label{sec:color}

In \cite{Guevara:2022qnm} we showed that the Jacobi identity of $w_{1+\infty}$
can be formulated as a particular multiresidue integral acting directly
on the massless graviton S-matrix.\footnote{This is also related to the residue analysis of \cite{Ball:2022vzc}. The main difference relies in that here and in \cite{Guevara:2022qnm} we assume a real contour along a null direction of the celestial torus.} This is precisely the kinematic
BCJ identity as originally formulated by Monteiro and O'Connell \cite{Monteiro:2014cda}. The
discussion will be now generalized to spinning-massive EFTs with minimal
gravitational coupling, which will connect with the double copy
constructions of the respective S-matrix.

In the following we will build 3-pt and 4-pt amplitudes by following
our colored CFT prescription. Let us first give a sketch of the argument, slightly extending the massless case to include matter fields. The main idea is to consider the following Kac-Moody algebra
\begin{align}
 O^{i}(z)O^{j}(z') & \sim \delta^{ij}\delta(z-z') + \frac{T^{ij}_{\,\,\,c}}{z-z'} \phi^c\\
G^{a}(z)O^{i}(z') & \sim\frac{T^{aij}}{z-z'}O_j(z') \\
G^{a}(z)G^{b}(z') & \sim\frac{if_{\quad c}^{ab}}{z-z'}G^{c}(z') \\
G^{a}(z)\phi^{b}(z') &\sim  \delta^{ab}\delta(z-z') + \frac{if^{ab}_{\,\,\,c}}{z-z'} \phi^c \label{eq:sdmss}
\end{align}
Here $O^i$ model matter fields and we assume $T^{ij}_a = -T^{ij}_a$. We can take $G^a$ to be a $SU(N)$ current playing the role of a $w_{1+\infty}$ graviton as $N\to \infty$, and $\phi^a$ to be a Goldstone mode. Because of the graviton-Goldstone paring in \eqref{eq:sdmss}, we can use this field to obtain 3-pt MHV amplitudes $\langle GG\phi\rangle\sim f^{abc}$, see \cite{Guevara:2022qnm} and discussion. From this algebra structure we can easily deduce the form of the three-point function\footnote{We normalize the delta function so that $2\delta(z)\leftrightarrow \frac{1}{z+i\epsilon}-\frac{1}{z-i\epsilon}$.}
\[
\langle G^{c}(z_{1})O^{i}(z_{2})O^{j}(z_{3})\rangle = T^{aij}\delta(z_{23})\delta(z_{12})
\]
We will associate the LHS to a three-point amplitude in a gravitational theory. To show that the above respects, say, the $GO$  OPE one needs to interpret the delta function $\delta(z_{12})$ as a pole, $\delta(z_{12}) \leftrightarrow \frac{1}{z_{12}}$ which is possible if the expression is to be integrated along the real line, corresponding to a null direction of the celestial torus. In section \ref{eq:spact} we will see this prescription is also consistent with the $(2,2)$ Penrose transform. At any rate, the integral results in
\begin{equation}
   \oint dz_2 dz_3 \langle G^{c}(z_{1})O^{i}(z_{2})O^{j}(z_{3})\rangle = T^{cij}  \,,\label{eq:3pti}
\end{equation}
and extracts the color structure $T^{cij}$ (in the fundamental representation) as a Jacobian. We will make extensive use of this idea: Our task is to simply extract a Jacobian that resolves the usual delta function singularities of scattering amplitudes. For instance, associativity of the algebra is encoded in the four-point function
\begin{equation}
    \langle G^{a}(z_{1}) G^{b}(z_{2})O^{i}(z_3) O^{j}(z_4)\rangle 
\end{equation}
The Jacobi identity is extracted from the three residues of the following contour integral 
\begin{equation}
    \oint dz_2 dz_3 dz_4 \langle G^{a}(z_{1}) G^{b}(z_{2})O^{i}(z_3) O^{j}(z_4)\rangle = T^{ai}_{\,\,\,k} T^{bkj}+ T^{bi}_{\,\,\,k} T^{ajk}+ if^{ab}_{\,\,\,e}T^{eij} =0 \label{eq:jaco}
\end{equation}
This is obtained by performing the multiresidue integral as a consecutive residue. Note that performing the $z_2$ integral reduces the expression to three integrals of the type \eqref{eq:3pti}.

Pushing forward this procedure, a natural conjecture emerges: 

\vspace{1em}

\textbf{Conjecture.} \textit{Consider a $n$-point function obtained via the celestial dictionary \eqref{eq:dcin} from a minimally coupled amplitude. We can employ a $n-1$-dimensional integral to extract color information satisfying associativity constraints. In particular, the integral
\begin{equation}
    \oint \prod_{i=2}^n dz_i  \langle G^{a_1}(z_{1})G^{a_2}(z_{2})   \cdots G^{a_{n-2}}(z_{n-2})O^{a_{n-1}}(z_{n-1}) O^{a_n}(z_n)\rangle 
\end{equation}
returns a color factor pertaining to a factorization channel. Algebraic relations between these color factors, such as \eqref{eq:jaco},  are then kinematic Jacobi identities in the S-matrix sense \cite{Bern:2019prr}. }
\vspace{1em}

While we leave a precise analysis of these integrals for future work, this result is expected from the analysis of \cite{Mizera:2019gea}, which addressed it from a closely related worldsheet perspective. Next we will give precise examples that, together with the massless cases studied in \cite{Guevara:2022qnm}, support this conjecture at $n=3,4$ points.

\subsection{OPE and 3-pt Amplitude}

Let us first construct the 3-point amplitude from the above kinematics.
We will start by adapting the collinear relation which emerges in
the minimal coupling, as defined by 

\begin{align}
G^{\eta_{1}}(z)O_{S}^{\lambda,\mu} & \sim\frac{(\epsilon\cdot p_{1})^{2}}{p\cdot k}\exp\left(\frac{F_{\mu\nu}J^{\mu\nu}}{2\epsilon\cdot p_{1}}\right)O_{S}^{\lambda,\mu}\\
 & =\frac{[\lambda1]}{z-\frac{[\mu1]}{[\lambda1]}}e^{\D_{\eta}}O_{S}^{\lambda,\mu}=\frac{[\lambda1]}{z-\frac{[\mu1]}{[\lambda1]}}O_{S}^{\lambda+\eta_{1},\mu+z\eta_{1}} \label{eq:opes}
\end{align}
Using this OPE we can easily obtain the 3-pt function,
\begin{align}
\langle G^{\eta_{1}}(z_1)O_{S}^{\lambda,\mu}O_{S}^{\lambda',\mu'}\rangle & =\frac{[\lambda1]}{z-\frac{[\mu1]}{[\lambda1]}}\langle O_{S}^{\lambda',\mu'}e^{\mathcal{D}_{\eta}}O_{S}^{\lambda,\mu}\rangle\\
 & =\frac{[\lambda1]}{z_1-\frac{[\mu1]}{[\lambda1]}}\langle\varepsilon'^{(k)}|e^{\mathcal{D}_{\eta}}|\varepsilon^{(j)}\rangle\,\delta^{2}(\lambda'+\lambda+\eta_{1})\delta\left(\frac{[\mu\xi]+z_1[1\xi]}{[\lambda\xi]+[1\xi]}-\frac{[\mu'\xi]}{[\lambda'\xi]}\right)
\end{align}
Recall that $\frac{[\mu'\xi]}{[\lambda'\xi]}=z'+\frac{[r_{\lambda'}'\xi]}{[\lambda'\xi]}$. Following the prescription \eqref{eq:3pti}, we can obtain the  color structure
\begin{align}
    \oint dz_1 dz' \langle G^{\eta_{1}}(z)O_{S}^{\lambda,\mu}O_{S}^{\lambda',\mu'}\rangle &= {[\lambda1]}\langle\varepsilon'^{(k)}|e^{\mathcal{D}_{\eta}}|\varepsilon^{(j)}\rangle\,\delta^{2}(\lambda'+\lambda+\eta_{1}) \label{eq:3min}\\
&=\langle\varepsilon'^{(k)}|T^{\lambda,\lambda',\eta_1}|\varepsilon^{(j)}\rangle \label{eq:sint}
\end{align}
In the second line we have implicitly defined the structure constants $T^{\lambda,\lambda',\eta_1}$ which are now operators acting on spin states. This generalizes the expression in \cite{Guevara:2022qnm}: The structure constant acquires an extra factor $e^{\mathcal{D}_{\eta}}$ accounting for internal space spin-rotation.

As anticipated, we can indeed go further and read off the minimal coupling three point amplitude off its residue. That is, we reinterpret
the 4D momentum conservation delta functions as holomorphic singularities
in the $z$ plane. In this case, this amounts to the replacement $\frac{1}{z-\frac{[\mu1]}{[\lambda1]}}\leftrightarrow\delta\left(z-\frac{[\mu1]}{[\lambda1]}\right)$.
Thus we get, up to a numerical factor
\begin{align}
\langle G^{\eta_{1}}(z)O_{S}^{\lambda,\mu}O_{S}^{\lambda',\mu'}\rangle & =[\lambda1]^{2}\langle\varepsilon'^{(k)}|e^{\mathcal{D}_{\eta}}|\varepsilon^{(j)}\rangle\,\delta^{2}(\lambda'+\lambda+\eta_{1})\delta\left(\frac{[\mu\xi]+z[1\xi]+[\mu'\xi]}{[\lambda'\xi]}\right)\nonumber\\
&\qquad \times \delta\left([\mu\lambda']+z[1\lambda']+[\mu'\lambda']\right)\nonumber \\
 & =(\epsilon\cdot p)^{2}\langle\varepsilon'^{(k)}|e^{\mathcal{D}_{\eta}}|\varepsilon^{(j)}\rangle\delta^{4}(p+p'+k)\label{eq:sdassx}
\end{align}
This form is valid for arbitrary spin $S$ and recovers the three-point
amplitude obtained in \cite{Guevara:2018wpp} for minimal coupling. In fact, the
massless limit is precisely the unique helicity-preserving three-point
amplitude. There is an underlying reason: Observe that the on-shell
boosted state $|\tilde{\varepsilon}^{(j)}\rangle:=e^{\D_{\eta}}|\varepsilon^{(j)}\rangle$
is nothing but a polarization vector associated to $p'=e^{\D_{\eta}}p$.
This implies that in the massless limit, the contraction
\begin{equation}
\langle\varepsilon'^{(k)}|e^{\D_{\eta}}|\varepsilon^{(j)}\rangle=\langle\varepsilon'^{(k)}|\tilde{\varepsilon}^{(j)}\rangle
\end{equation}
is both gauge invariant for $p,p'$ and Lorentz invariant. As there
is a unique three-point amplitude fixed by helicity weights and Lorentz
invariance, we conclude that (\ref{eq:sdassx}) is the correct massless
amplitude. Since it is written in covariant language, it is natural
to promote it to a massive amplitude without adding $m^{2}$ deformations:
This is precisely the definition of minimal coupling given in \cite{Arkani-Hamed:2017jhn}
(see also \cite{Bautista:2019evw}).

Regarding the relation to the massless case, it is worth mentioning that the massless limit of (\ref{eq:sdassx})
unifies all sectors, i.e. the 8 helicity configurations $(\pm,\pm,\pm)$.
In particular one can show that the same helicity amplitudes $(+++)$
vanish as expected for minimal coupling.

\subsection{Jacobi Identity and 4-pt Amplitude}

The main observation can be introduced at $n=4$ points. Consider
two incoming gravitons and two spin-$S$ matter insertions
\begin{equation}
\langle G^{\eta_{1}}(z_{1})G^{\eta_{2}}(z_{2})O_{S}^{\lambda'}(z')O_{S}^{\lambda}(z)\rangle 
\end{equation}
This amplitude corresponds to gravitational Compton scattering and was derived from the minimal coupling criteria \cite{Arkani-Hamed:2017jhn,Johansson:2019dnu,Bautista:2019evw}. Its explicit form is not needed here (but can be easily seen to agree with equation \eqref{eq:sx1m} below).

We will repeat the massless case computation \eqref{eq:jaco} and show that it spills out nothing but the  BCJ kinematic numerators. Our goal is to extract the residue 
\begin{equation}
\oint dz'\,dz_{1}dz_{2}\langle G^{\eta_{1}}(z_{1})G^{\eta_{2}}(z_{2})O_{S}^{\lambda',r'}(z')O_{S}^{\lambda,r}(z)\rangle
\end{equation}
over all codimension-3 singularities in $(z,z_{1},z_{2})$.  Let us compute the three residues one by one. Start with the singularity as $z\to z_1$.
\begin{equation}
    \langle G^{\eta_{1}}(z_{1})G^{\eta_{2}}(z_{2})O_{S}^{\lambda',\mu'}(z')O_{S}^{\lambda,\mu}(z)\rangle  \to \frac{[\lambda 1]}{z_1-[\mu 1]/[\lambda 1]}   \langle G^{\eta_{2}}(z_{2})O_{S}^{\lambda',\mu'}(z')e^{\hat{\mathcal{D}}_{\eta_1}}O_{S}^{\lambda,\mu}(z)\rangle 
\end{equation}
We refer to this factorization channel as s channel. Using \eqref{eq:3min} we obtain
\begin{align}
      \oint_s dz' dz_1 dz_2 \langle G^{\eta_{1}}(z_{1})G^{\eta_{2}}(z_{2})O_{S}^{\lambda',\mu'}(z')O_{S}^{\lambda,\mu}(z)\rangle &= [\lambda 1 ][\lambda'2] \langle \varepsilon'^{(k)}|e^{\mathcal{D}_{\eta_2}}e^{\mathcal{D}_{\eta_1}}|\varepsilon^{(j)}\rangle \delta^{2}(\eta_1 +\eta_2 +\lambda +\lambda') \label{eq:sres}\\
&=\langle\varepsilon'^{(k)}|T^{\lambda,\rho, \eta_1 } T^{\lambda',\rho,\eta_2 }|\varepsilon^{(j)}\rangle
\end{align}
where the sum over $\rho_\alpha$, associated to the internal massive state, is implicit. This is indeed equivalent to a contraction of two copies of color structures \eqref{eq:3min} using the operator product to multiply the spin factors $e^{\mathcal{D}_{\eta}}$. We can obtain the $u$ channel similarly by exchanging $1\leftrightarrow 2$. Here we have used that on the support of the poles in $z_1,z_2$ the on-shell operators $\D_{\eta}$ can be replaced by the displacements $\mathcal{D}_\eta$, which commute. The last residue is obtained as $z_1 \to z_2$ and thus illustrates the symmetry algebra $Lw_{1+\infty}$  relation \eqref{eq:womf}:
\begin{equation}
        \langle G^{\eta_{1}}(z_{1})G^{\eta_{2}}(z_{2})O_{S}^{\lambda',\mu'}(z')O_{S}^{\lambda,\mu}(z)\rangle  \to \frac{[12]}{z_1-z_2}   \langle G^{\eta_1+\eta_{2}}(z_{2})O_{S}^{\lambda',\mu'}(z')O_{S}^{\lambda,\mu}(z)\rangle 
\end{equation}
Again using  \eqref{eq:3min}  yields
\begin{align}
       \oint_t dz' dz_1 dz_2 \langle G^{\eta_{1}}(z_{1})G^{\eta_{2}}(z_{2})O_{S}^{\lambda',\mu'}(z')O_{S}^{\lambda,\mu}(z)\rangle &= [12][\lambda \lambda'] \langle \varepsilon'^{(k)}|e^{\D_{\eta_1+\eta_2}}|\varepsilon^{(j)}\rangle \delta^{2}(\eta_1 +\eta_2 +\lambda +\lambda') \label{eq:tsm2} \\
       &=if^{\eta_1 \eta_2 \rho}  \langle\varepsilon'^{(k)}|T^{\lambda \lambda'\rho} |\varepsilon^{(j)}\rangle\,.
\end{align}
Here
\begin{equation}
    f^{\eta_1 \eta_2 \rho} := [12] \delta^2(\eta_1+\eta_2+\rho) \label{eq:antse}
\end{equation}
is totally antisymmetric and matches the definition in \cite{Guevara:2022qnm}. Note that on the support of $z_1=z_2$ we can again replace
\begin{equation}
\D_{\eta_1+\eta_2}\to\mathcal{D}_{\eta_1}+\mathcal{D}_{\eta_2}\,,\end{equation}
which commute with each other. It then follows that the color factor depending on polarization tensors can be matched among $s,t$ and $u$ channels. A very direct computation then shows that
\begin{equation}
     \oint_{s+t+u} dz' dz_1 dz_2 \langle G^{\eta_{1}}(z_{1})G^{\eta_{2}}(z_{2})O_{S}^{\lambda',\mu'}(z')O_{S}^{\lambda,\mu}(z)\rangle =0 \label{eq:tres}
\end{equation}
where we have added up the three residues. 

Let us now discuss the interpretation of the above in terms of color kinematics duality. For the two massive-two graviton amplitude, the BCJ decomposition has been worked out in \cite{Johansson:2019dnu,Bautista:2019tdr,Bautista:2019evw}:
\begin{equation}
\langle G^{\eta_{1}}(z_{1})G^{\eta_{2}}(z_{2})O_{S}^{\lambda',r'}(z')O_{S'}^{\lambda,r}(z)\rangle=\left(\frac{n_{t}\tilde{n}_{t}}{t}+\frac{n_{s}\tilde{n}_{s}}{s-m^{2}}+\frac{n_{u}\tilde{n}_{u}}{u-m^{2}}\right)\delta^2(\mu+\mu'+z_1\eta_1+z_2\eta_2) \label{eq:sx1m}
\end{equation}
Here the numerators $\tilde{n}_i$ carry the spin information, where we have absorbed two delta functions. The $n_i$ are scalar numerators.  Not surprisingly, the spin numerators are given by \eqref{eq:sres} and \eqref{eq:tsm2}. Our results are then equivalent to the identity
\begin{equation}
    \oint dz'\,dz_{1}dz_{2}\, \frac{n_i}{P_i} \times \delta^2(\mu+\mu'+z_1\eta_1+z_2\eta_2)  =1 \label{eq:cknd}
\end{equation}
where $n_i \in \{n_s,n_u,n_t\}$ and $P_i\in \{t,s-m^2,u-m^2\}$. This in turn yields, when applied to the sum of the channels:
\begin{align}
    \tilde{n}_t+\tilde{n}_s+\tilde{n}_u=0
\end{align}
This is indeed the kinematic Jacobi identity, which is the defining signature of color kinematics duality. Let us show the identity \eqref{eq:cknd} explicitly for the s-channel, with the other channels proceeding analogously. We have
\begin{equation}
    \frac{n_s}{s-m^2}=\frac{[\lambda 1][\lambda'2]}{[\lambda 1](z_1-z-[r_{\lambda}1]/[\lambda 1])} = \frac{[\lambda'2]}{(z_1-z-[r_{\lambda}1]/[\lambda 1])}\label{eq:nsnd}
\end{equation}
We will use the contour around this pole to fix the $z$ variable. Then 
\begin{equation}
    \delta^2((z_1 -[r_{\lambda}1]/[\lambda 1]) \lambda+r_\lambda +\mu'+z_1\eta_1+z_2\eta_2) = \delta^2(z_1 (\lambda+\eta_1)+z_2\eta_2+\ldots)
\end{equation}
Hence $z_1$ and $z_2$  can be integrated to impose momentum conservation, yielding a Jacobian:
\begin{equation}
    \delta^2(z_1 (\lambda+\eta_1)+z_2\eta_2+\ldots)= \frac{1}{[\lambda+\eta_1,\eta_2]}\delta(z_1-\cdots) \delta(z_2-\cdots)
\end{equation}
which cancels the numerator of \eqref{eq:nsnd} on the support of the delta function inside $\tilde{n}_s$. Putting all together, equation \eqref{eq:cknd} follows. Thus we have shown that the kinematic algebra is equivalent to the residue condition \eqref{eq:tres}.

\section{Beyond the wedge: Off-shell $w_{1+\infty}$ algebra }\label{eq:offshells}
Thus far we have discussed the on-shell residues of the EFT amplitudes, and found that the off-shell operators $\mathcal{D}_\eta$ commute in the support of the cuts. However, in defining the OPE and the algebra in CCFT, we require an on-shell extension. This is because in extracting commutators we get 
\begin{align}
    \oint dz_1 G^{\eta}(z_1) O^{\lambda,\mu}_S&= [G^\eta, O^{\lambda,\mu}_S]\\
    &= [\lambda \eta] e^{\hat{\mathcal{D}}_\eta} O^{\lambda,\mu}_S \label{eq:cmt}
\end{align}
where the on-shell generator is
\begin{equation}
    \hat{\mathcal{D}}_\eta = \frac{[\eta|J|\eta]}{[\lambda \eta]}
\end{equation}
We can now ask if this action satisfies a version of $w_{1+\infty}$ algebra. In particular,
\begin{equation}
     [G^{\eta_1}, [G^{\eta_2},O^{\lambda,\mu}_S] -     [G^{\eta_2}, [G^{\eta_1},O^{\lambda,\mu}_S] \stackbin{?}{=} [12][G^{\eta_1+\eta_2},O^{\lambda,\mu}_S] \label{eq:amns}
\end{equation}
is true if
\begin{equation}
    e^{\mathcal{D}_{\eta_1}}e^{\mathcal{D}_{\eta_2}} \stackbin{?}{=}e^{\mathcal{D}_{\eta_1+\eta_2}} \label{eq:sa}
\end{equation}
The key property of $\hat{\mathcal{D}}_\eta$ as opposed to its off-shell version $\mathcal{D}_\eta$ is that 1) it is non-linear in $\eta$, the massless momentum and 2) generators associated to different momenta do not commute. The first observation implies that the expansion $|\eta] \to \omega (\bar{z}|+]+|-])$  does not truncate in $\bar{z}$ and hence generate all operators beyond what is called the wedge algebra \cite{Guevara:2021abz}. The second observation comes from analyzing the commutator
\begin{equation}
  [\mathcal{D}_{\eta_1},\mathcal{D}_{\eta_2}] = \left[ \frac{[1|J|1]}{[\lambda 1]}, \frac{[2|J|2]}{[\lambda 2]}\right ] = \frac{[12]^3}{[1\lambda]^2[2\lambda]^2} [\lambda|J|\lambda] \label{eq:mss2}
\end{equation}
We also have 
\begin{equation}
    \mathcal{D}_{\eta_1}+\mathcal{D}_{\eta_1}-\mathcal{D}_{\eta_3} = \frac{[12]^2}{[1\lambda][2\lambda][3\lambda]} [\lambda|J|\lambda] \label{eq:mss3}
\end{equation}
with $\eta_3=\eta_1+\eta_2$. These two equations suggests a way to recover the algebra relation \eqref{eq:amns}. Observe that the desired algebra \eqref{eq:amns} is linear in $[12]$ while \eqref{eq:mss2} and \eqref{eq:mss3} are at least quadratic. Hence, by virtue of the Baker-Campbell-Hausdorff formula we recover \eqref{eq:sa} by dropping terms $\mathcal{O}([12]^2)$ which can be interpreted as quantum corrections.\footnote{Note that the power of $[12]$ counts how many times a commutator of the $SL(2,\mathbb{R})$ algebra has been applied. For instance, in computing \eqref{eq:mss2} we need to commute $[1|J|1]\leftrightarrow [\lambda 2], [2|J|2]\leftrightarrow[\lambda 1]$ and $[1|J|1]\leftrightarrow [2|J|2]$. In a CFT, each use of the commutator amounts for a Wick contraction.} Indeed this is what happens in the massless case since 
\begin{equation}
    c=[\lambda|J|\lambda]=[\lambda\mu][\lambda\frac{\partial}{\partial \mu}] = m\partial_Z
\end{equation}
vanishes. We can denote $\bar{L}_{-}=\partial_Z $ (see \eqref{eq:pffm}), which represents the action of a `left' Lorentz group, i.e. commutes with $J$. Because of this, the combination is indeed a $c$-number that parametrizes the deformation of the $w_{1+\infty}$ algebra, as we detail next.

\subsection{Towards a quantum $w_{1+\infty}$ algebra}
Our results above imply that
\begin{equation}
    [G^\eta,\cdot ]= [\eta \lambda] e^{\D_\eta}
\end{equation}
satisfies the $w_{1+\infty}$ algebra up to powers of $[\lambda|J|\lambda]$. For comparison with the literature we can extract the modes of $G^\eta$ which correspond to independent generators of the algebra. This is easier by introducing the energy parameter $\eta=\omega \hat{\eta}$, leading to the expansion
\begin{equation}
    G^\eta= \sum_k \frac{1}{k!} [\eta \lambda]\D^k_{\eta}
\end{equation}
where the order $\omega^{k+1}$ is given by
\begin{equation}
    W^{\frac{k+3}{2}}_\eta = [\eta \lambda]^{1-k}[\eta|J|\eta]^k
\end{equation}
i.e. $W^{3/2}=[\eta \lambda],W^2=[\eta|J|\eta]$, etc... These generators belong to the universal enveloping algebra of $[\eta\lambda]$ and $[\eta|J|\eta]$. We will now show how to construct it.\footnote{Similar constructions based on quotients of enveloping algebras appeared in \cite{Bittleston:2023abc,Bogna:2024abc}, in the context of algebras on self-dual spacetimes.} Given two massless directions $\eta_1$ and $\eta_2$ ,we define
\begin{align}
L_-=[1|J|1]\,,\,L_+=&[2|J|2]\,,\,L_0= [1|J|2]\\
\lambda_{-1/2} = [\lambda 1]\,,\,\, & \lambda_{+1/2} = [\lambda 2]
\end{align}
It follows from \eqref{eq:prealg} that they satisfy
\begin{align}
    [L_m,L_n]&=[12](m-n)L_{m+n} \\
    [L_m,\lambda_{n}]&=[12] (\frac{3}{2}m - n)\lambda_{m+n}\\
    [\lambda_m,\lambda_n]&=0\,.
\end{align}
It is well known that the enveloping algebra of $\{L_n\}$, namely $U(sl(2,\mathbb{R}))$  can be used to reconstruct a one-parameter family $w_{\infty}(L^2)$, where we quotient by the Casimir 
\begin{equation}
   L^2 =  \frac{L_+ L_- + L_- L_+}{2} + L_0^2 \label{eq:sadx}
\end{equation}
In contrast, in our case we are interested in the envelope of $\{L_m,\lambda_n\}$ which adds half-integer modes. In this case \eqref{eq:sadx} is not a Casimir of the algebra, since we expect $[L^2,\lambda_m] = \lambda_m$ in the adjoint representation. Instead, the following combination
\begin{equation}
    [12] c =  \lambda_-^2 L_+ + 2\lambda_+ \lambda_-  L_0  + \lambda_+^2 L_{-}
\end{equation}
is a Casimir and leads to a 1-parameter deformation $w_{1+\infty}(c)$. Denoting by $\mathcal{D}_+ = \mathcal{D}_{\eta_1}=L_{+}/\lambda_{+}$ and $\mathcal{D}_-=\mathcal{D}_{\eta_2}=L_{-}/\lambda_{-}$ we have
\begin{equation}
    W^{\frac{k+3}{2}}_{\alpha_1\cdots \alpha_{k+1}} = \lambda_{(\alpha_1}\mathcal{D}_{\alpha_2}\cdots \mathcal{D}_{\alpha_{k+1})} \label{eq:paralas}
\end{equation}
Noting that $[\mathcal{D}_{\alpha},\lambda_{\beta}]=[12]\epsilon_{\alpha\beta}$  and that $[\mathcal{D}_{\alpha},\mathcal{D}_{\beta}]=\mathcal{O}(c)$ (according to \eqref{eq:mss2}) we immediately have
\begin{equation}
    [ W^{\frac{k+3}{2}}_{\alpha_1\cdots \alpha_{k+1}}, W^{\frac{j+3}{2}}_{\beta_1\cdots \beta_{j+1}}] = [12]\epsilon_{\alpha_1 \beta_1} W_{\alpha_2 \cdots \alpha_{k+1} \beta_2 \cdots \beta_{j+1}} + \mathcal{O}(c)\,.\label{eq:sss1}
\end{equation}
This realizes the commutator \eqref{eq:amns} in this language. Recall that in the massless case $c=m\bar{L}_{-1}=0$ and this is exactly the $w_{1+\infty}$ algebra.

The wedge algebra now follows from the identity \eqref{eq:mss3}, by setting $|3]=|\eta]=\bar{z}|1]+|2]$ we find that 
\begin{equation}
    W_{\eta}^{\frac{k+3}{2}}= W^{\frac{k+3}{2}}_{\alpha_1\cdots \alpha_{k+1}} \eta^{\alpha_1}\cdots \eta^{\alpha_{k+1}} +\mathcal{O}(c)\label{eq:sss3}
\end{equation}
where the first term is polynomial in $\bar{z}$ and belongs to the wedge. Hence $c$ also parametrizes `massive' deviations from the wedge algebra. We leave the precise computations of $c$- terms in \eqref{eq:sss1} and \eqref{eq:sss3} to future work.

\section{ Classical Spacetimes from Color Structure}\label{eq:spact}

It has recently been appreciated that the EFT approach allows us to relate scattering amplitudes and perturbative solutions of general relativity. The situation is particularly useful for rotating black holes. This is because their metric and curvature follow from a $S\to \infty$ limit of amplitudes of a spinning massive particle coupled to graviton. This connection provides a new meaning to the color structure constants identified in equation \eqref{eq:3min}.

Here we will sketch the connection and leave a more detailed treatment for future work. We will follow heavily the computation of \cite{Guevara:2021abx}, which we refer for more details on the integrals performed here.

According to the KMO prescription \cite{Kosower:2018adc}, to derive a scattering amplitude corresponding to a classical system we first need to integrate over the phase space of the outgoing massive particle. In our notation
\begin{equation}
   \langle \mathcal{M}_3 (k)\rangle = \int d^4p' \delta(p'^2-m^2) \mathcal{M}_3 (p,p',k_1) = \int dz' d^2\lambda'  \langle G^{\eta}(z_1)O_{S}^{\lambda,\mu}O_{S}^{\lambda',\mu'}\rangle
\end{equation}
Now, the classical amplitude can be half-Fourier transformed to obtain a \textit{twistor integrand} $f_{-2}(Z)$, a meromorphic function in $Z=(1,z_1,\kappa_+,\kappa_-)\in\mathbb{CP}^3$:
\begin{equation}
    f_{-2}(z_1,\kappa)= \int d^2\eta e^{[\eta \kappa]} \langle \mathcal{M}_3(k)\rangle 
\end{equation}
where $k=|\eta\rangle [\eta|$ and $|\eta\rangle = (1,z_1)$. To obtain the spacetime curvature, we then set $|\kappa]=x|\eta\rangle$ (the incidence relation) and evaluate the \textit{Penrose transform}:
\begin{equation}
    C_{++++}= \oint dz_1 f_{-2} (z_1, x|\eta\rangle)  
\end{equation}
where $C_{++++}$ denotes the top component of the spacetime Weyl tensor. We will not elaborate here on contour of the transform, which is determined by the cohomology of twistor space. Rather, we simply follow the prescription of the previous section and treat poles in $z_1$ and delta functions interchangeably. With this in mind, we can finally write
\begin{align}
        C_{++++} &=\int d^2\eta d^2\lambda' e^{[\eta |x|\eta\rangle} \oint dz_1 dz' \langle G^{\eta}(z_1)O_{S}^{\lambda,\mu}O_{S}^{\lambda',\mu'}\rangle \\
       &= \int d^2\eta d^2\lambda' e^{\frac{[\eta |x \hat{p}|\eta]}{[\eta \lambda]}} \langle\varepsilon'^{(k)}|T^{\lambda,\lambda',\eta}|\varepsilon^{(j)}\rangle \label{eq:cpp}
\end{align} 
In this formula we have used that the delta function support fixes, according to the parametrization of section \ref{sec:kin},
\begin{equation}
    |\eta\rangle= (1 \,\,,\,\,z_1) = \frac{1}{[\lambda \eta]}( [\lambda \eta] \,\,,\,\,[\lambda \mu])= \hat{p}|\eta]/[\eta\lambda]
\end{equation}
where $\hat{p}=p/m$. Following \cite{Guevara:2021abx}, the combination $(x\hat{p})_{\alpha\beta}$ can be understood as the orbital angular momentum of spacetime. 

Our result \eqref{eq:cpp}  establishes a direct connection between the color structures pertaining a Kac-Moody CFT and a classical solution in general relativity. Furthermore,  it provides a different justification for the integration prescription \eqref{eq:sint}, namely the suitable integration contour is dictated by the Penrose transform. 

We can evaluate the formula explicitly for the case of spinning minimally coupled particle. The classical limit is obtained by removing the spin indices \cite{Guevara:2018wpp}
\begin{equation}
\langle\varepsilon'^{(k)}|T^{\lambda,\lambda',\eta} |\varepsilon^{(j)}\rangle\to T^{\lambda,\lambda',\eta} = {[\lambda\eta]} e^{\frac{[\eta |J|\eta]}{[\eta \lambda]}} \,\delta^{2}(\lambda'+\lambda+\eta_{1})\label{eq:repsdl}
\end{equation}
and interpreting the operator $J$ on the RHS as a c-number, the spin parameter of the spacetime. Denoting $\mathcal{L}(x)=x\hat{p}+J$ and plugging this into \eqref{eq:cpp} gives
\begin{align}
   C_{++++} &= \int d^2\eta [\lambda\eta] e^{\frac{[\eta |\mathcal{L}(x)|\eta]}{[\eta \lambda]}}\\
   & = \oint [\eta d\eta]\frac{[\lambda\eta]^4 }{[\eta |\mathcal{L}(x)|\eta]^3}
\end{align}

In the second line, we have integrated out the energy scale $\omega$ (by setting $\eta\to \omega \eta$). The result is a projective integral, again of the Penrose form. This formula can be matched to the Kerr black hole metric, where we can decompose the Killing spinor as $\mathcal{L}_{\alpha\beta}(x)=\omicron_{(\alpha}\iota_{\beta)}$ \cite{Guevara:2021abx}. Then $\omicron_{\alpha}(x)$ and $\iota_{\alpha}(x)$ are called Principal Null Directions (PNDs) of the black hole spacetime.

We are left to discuss the four-point amplitude. We can use expression \eqref{eq:cmt} to obtain
\begin{equation}
    \langle [G^{\eta_1},G^{\eta_2}(z_2)] O_S^{\lambda,\mu} O_S^{\lambda',\mu'}\rangle= \oint_t dz_1 \langle G^{\eta_1}(z_1) G^{\eta_2}(z_2) O_S^{\lambda,\mu} O_S^{\lambda',\mu'}\rangle
\end{equation}
where the $t$ residue amounts to $z_1\to z_2$. The LHS can be interpreted simply as the action of a $w_{1+\infty}$  generator: $[G^\eta,\cdot]=\delta_{\eta}(\cdot)$. Combining this with \eqref{eq:cpp} then implies
\begin{align}
        \delta_{\eta_1}C_{++++} &=\int d^2\eta_2 d^2\lambda' e^{[\eta_2 |x|\eta\rangle} \oint dz_1 dz_2 dz' \langle G^{\eta_1}(z_1)G^{\eta_2}(z_2)O_{S}^{\lambda,\mu}O_{S}^{\lambda',\mu'}\rangle \\
       &= \int d^2\eta_2 d^2\lambda' e^{\frac{[\eta_2 |x \hat{p}|\eta_1+\eta_2]}{[\eta_1+\eta_2, \lambda]}} if^{\eta_1,\eta_2,\rho}\langle\varepsilon'^{(k)}|T^{\lambda,\lambda',\rho}|\varepsilon^{(j)}\rangle \label{eq:cp2te}
\end{align} 
Again, the contour prescription proposed in the previous section takes the form of a Penrose transform. As an example, we can evaluate this formula at linear order in $\eta_1$. From \eqref{eq:paralas} we expect this to be $W^{3/2}_{\alpha}$. We obtain, recalling \eqref{eq:antse},
\begin{equation}
    [W^{3/2}_\alpha,C_{++++}] =  = \int d^2\eta d^2\lambda' e^{\frac{[\eta |x \hat{p}|\eta]}{[\eta \lambda]}}\eta_{\alpha} \langle\varepsilon'^{(k)}|T^{\lambda,\lambda',\eta}|\varepsilon^{(j)}\rangle \label{eq:csadss}
\end{equation}
By direct comparison with \eqref{eq:cpp}, the above is equivalent to $\partial_{+\alpha}C_{++++}$, illustrating that the action of $W^{3/2}$ on the Kerr metric is simply a translation as expected. The action of higher generators $W^{k}$, $k>3/2$, which follows from \eqref{eq:cp2te}, appears more complicated and we leave a complete treatment for future work.

\section{Discussion}

In this work we have continued to outline new connections between color-kinematic duality, associativity of celestial CFT, and classical solutions in General Relativity. These lines of research are vast, and we expect that the ideas here can be generalized. In particular, we expect our conjecture at the end of section \ref{sec:color} to hold based on results on color-kinematics and factorization \cite{Mizera:2019gea,Bautista:2019tdr}.

We would like to conclude by pointing out two interesting directions that we believe deserve futher exploration.

\subsubsection*{Future direction: Color Kinematics in Conformal Basis}

Celestial holography is often formulated in the language of boost eigenstates obtained via Mellin transformation. While there are some subtleties in the dictionary for massive states \cite{Himwich:2023wso}, a direct translation of the OPEs we discussed can be done for the graviton states. For instance, consider the OPE
\begin{equation}
    G^{\eta_1}(z_1) G^{\eta_2}(z_2) \sim \int d^2\eta_3 \,\frac{f^{\eta_1,\eta_2}_{\,\,\quad \eta_3}}{z_{12}} G^{\eta_3}(z_2)\label{eq:sope}
\end{equation}
where $f^{\eta_1,\eta_2}_{\,\,\quad \eta_3}$ is our $SU(N\to\infty)$ structure constant and we have made explicit the summation convention. This expression can be easily translated to the OPE block for gravity including all conformal descendants. Seting $\eta_i = \omega_i(1\,\,\,\bar{z}_i)$ and introducing
\begin{equation}
    f^{\Delta_1,\Delta_2}_{\,\quad\,\Delta_3}(\bar{z}_i) = \int \prod d\omega_i \omega_i^{\Delta_i
  -1 } f^{\eta_1,\eta_2}_{\,\,\quad \eta_3}\,,
\end{equation}
one finds that \eqref{eq:sope} can be written as the chiral OPE block of \cite{Guevara:2021abz}
\begin{equation}
    G^{\Delta_1}(z_1,\bar{z}_1) G^{\Delta_2}(z_2,\bar{z}_2) \sim \int d\Delta_3 d\bar{z}_3 \,\frac{ f^{\Delta_1,\Delta_2}_{\,\quad\,\Delta_3}(\bar{z}_i)}{z_{12}} G^{\Delta_3}(z_2,\bar{z}_3)
\end{equation}
Thus the conformal basis also can be seen as a change of basis acting directly on the structure constants, and the Jacobi/color kinematics identity proceeds as before. It would be interesting to generalize this for the massive  states.

\subsubsection*{Future direction: Wilson lines}

Our results are closely related to the gravitational Wilson lines, used in \cite{Crawley:2023brz} to describe a BH state. In that paper a state $|\psi\rangle_a$ for the classical rotating black hole was constructed by inserting a Wilson line defect in the spacetime bulk, become a coherent state in the 2D CFT. The key formula used there, in our notation, reads 
\begin{equation}
   \,_a \langle \psi | G^{\eta}(z)|\psi\rangle_a = (\epsilon \cdot p)^2 e^{a\cdot k} \delta(k\cdot p) \propto [\eta \lambda ] e^{\frac{[\eta|J|\eta]}{[\lambda \eta]} }\delta\left(z-\frac{[\mu\eta]}{[\lambda \eta]}\right) \label{eq:rsc}
\end{equation}
where we have used $J_{\mu\nu}\leftrightarrow p_{[\mu}a_{\nu]}$ and $p_\mu$ is the time direction/mass of the spacetime, following the parametrization \eqref{eq:p1fix}.  Comparison with the OPE \eqref{eq:opes} immediately suggests the relation
\begin{equation}
    |\psi\rangle_a \leftrightarrow O^{\lambda,\mu} \label{eq:whzz}
\end{equation}
However, acting on the RHS the operator $\D_{\eta}$ is noncommutative as opposed to the operator $J$ in \eqref{eq:rsc}. In that sense, the massive operator $O^{\lambda,\mu}$ represents a quantum noncommutative correction to the state $|\psi\rangle_a$ which may become relevant with the insertion of two or more gravitons. 

Conversely, this suggests a Wilson line representation for $O^{\lambda,\mu}$. Indeed, the results of \cite{Crawley:2023brz} can be formulated in terms of the dual field \eqref{eq:csadss} which plays the role of a negative helicity state. We have
\begin{equation}
    |\psi\rangle_a= \exp\left(\int d^2\eta\, d^2 \lambda' T^{\lambda ,\lambda' ,\eta} \phi^{\eta} ([\mu \eta]/[\lambda \eta])\right)|0\rangle 
\end{equation}
which mimics a group element $e^{i T^c \phi_c}$ in a color algebra. This suggests that the massive operators shall be identified with the Wilson line
\begin{equation}
    O^{\lambda,\mu} = \exp\left(\int d^2\eta\, d^2 \lambda' \hat{T}^{\lambda ,\lambda' ,\eta} \phi^{\eta} ([\mu \eta]/[\lambda \eta])\right) \hat{O}^{\lambda ,\mu}
\end{equation}
where $\hat{O}^{\lambda ,\mu}$ is an operator that does not interact with gravity. Here we have replaced $T^{\lambda ,\lambda' ,\eta}$ by its noncommutative version $\hat{T}^{\lambda ,\lambda' ,\eta}$. This representation makes explicit the non-local nature of the massive operators in regards to the OPE, which certainly deserves further exploration. At the same time, similar representations have recently played a role in formulating a quantum error correcting code in this context \cite{Guevara:2023tnf,GHcode}. Such analysis also relies on the symmetry structure and it is likely possible to extend it to the massive case.

\subsubsection*{Acknowledgements}

We would like to thank Adam Tropper, Clifford Cheung, Henrik Johansson, Ricardo Monteiro and Oliver Schlotterer for useful conversations. We thank Mina Himwich and Monica Pate for preliminarly sharing some of the results in \cite{Himwich:2023wso}. We further wish to thank the Isaac Newton Institute, Cambridge UK, for support and hospitality during the programme `Twistor Theory' where work on this paper was completed.  We gratefully acknowledge support from the Black Hole Initiative and the Society of Fellows at Harvard University, as well as the DOE grant DE-SC0007870 and EPSRC grant no EP/R014604.

\bibliographystyle{utphys}
\bibliography{reference}

\end{document}